\documentclass{article}
\usepackage{spconf,amsmath,amsfonts,graphicx,multirow,hyperref,amssymb,algpseudocode,algorithmicx,algorithm,xcolor,booktabs,amsthm}

\usepackage[font=small,labelfont=bf]{caption} % 标题更紧凑
\usepackage{url}

\title{Holographic Transformers for Complex-Valued Signal Processing: Integrating Phase Interference into Self-Attention}
\name{
\begin{tabular}{c}
Enhao Huang$^{1}$, Zhiyu Zhang$^{1}$, Tianxiang Xu$^{2}$, Chunshu Xia$^{1}$, Kaichun Hu$^{1}$\\
Yuchen Yang$^{1}$, Tongtong Pan$^{1}$, Dong Dong$^{1}$, Zhan Qin$^{1}$\sthanks{Corresponding author.}
\end{tabular}
}
\address{
$^{1}$ College of Computer Science and Technology, Zhejiang University, China \\ 
$^{2}$ School of Software and Microelectronics, Peking University, China 
}

\begin{document}

\ninept

\maketitle

\begin{abstract}
Complex-valued signals encode both amplitude and phase, yet most deep models treat attention as real-valued correlation, overlooking interference effects. We introduce the Holographic Transformer, a physics-inspired architecture that incorporates wave interference principles into self-attention. Holographic attention modulates interactions by relative phase and coherently superimposes values, ensuring consistency between amplitude and phase. A dual-headed decoder simultaneously reconstructs the input and predicts task outputs, preventing phase collapse when losses prioritize magnitude over phase. We demonstrate that holographic attention implements a discrete interference operator and maintains phase consistency under linear mixing. Experiments on PolSAR image classification and wireless channel prediction show strong performance, achieving high classification accuracy and F1 scores, low regression error, and increased robustness to phase perturbations. These results highlight that enforcing physical consistency in attention leads to generalizable improvements in complex-valued learning and provides a unified, physics-based framework for coherent signal modeling. The code is available at \url{https://github.com/EonHao/Holographic-Transformers}.

\end{abstract}


\section{Introduction}\label{sec:intro}

Complex-valued signal representations are widely used across science and engineering due to their ability to jointly encode both amplitude and phase information. Amplitude typically reflects energy or reflectivity (e.g., backscatter strength in PolSAR), while phase captures structural details and coherence patterns that are essential for effective analysis and inference. Exploiting both components is critical for robust performance in domains such as audio, radar, communications, and biomedical signals \cite{Bassey2021Survey,Trabelsi2018DeepComplex,Savadkoohi2020SeizurePrediction,Fuchs2021Radar}.

Deep learning has made significant inroads into the complex domain, with convolutional and recurrent architectures achieving substantial improvements in tasks like speech enhancement and waveform restoration \cite{Choi2019DCUNet,Hu2020DCCRN}. However, these models focus on local context and often treat phase as secondary, either by emphasizing magnitude losses or decoupling the real and imaginary components. This causes fragility when phase distortions or long-range dependencies arise \cite{Takahashi2018Phasenet,cai2020phasednn,venkatasubramanian2025steinmetz}. This limitation is particularly evident in wave-like data, where long-range interactions like interference, dispersion, and coherent summation govern the observed signal structure \cite{tay2021lra}.

Transformers, with their ability to model long-range dependencies through self-attention, have been increasingly applied to complex-valued signal processing tasks \cite{Vaswani2017Attention,Yang2020ComplexTransformer}. While these approaches have shown improvements, they still face limitations. Many methods reduce complex features to magnitudes or split real and imaginary parts \cite{Yin2020PHASEN}, while others compute attention using real-valued similarities even for complex tokens \cite{Peng2024SignalTransformer,Eilers2023ComplexTransformerBlocks}. These strategies fail to capture key mechanisms of coherent superposition, such as constructive and destructive interference, leading to representations misaligned with the underlying physics. This limitation arises because phase information is often reduced to amplitude values, neglecting the physical properties inherent in phase \cite{Dramsch2021PhaseMatters}. Traditional deep learning models, including attention mechanisms, are designed for magnitude-based representations \cite{Wu2023PhaseMatters}, focusing on amplitude correlations, which fails to capture the complexities of phase interactions in complex-valued signals.

\begin{figure}[t]
    \centering
    \includegraphics[width=\linewidth, trim=6 6 6 6, clip]{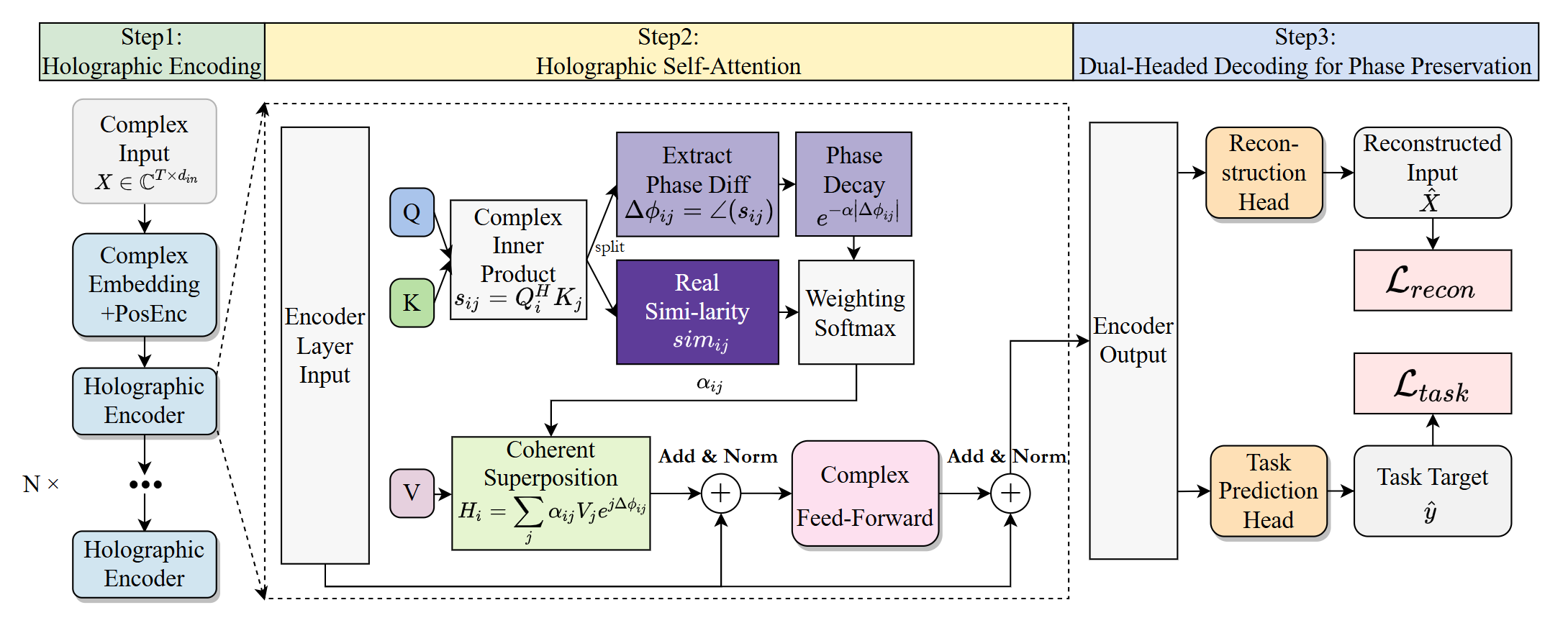}
    \caption{Holographic Transformer Architecture}
    \vspace{-0.3em}
    \label{fig:label}
\end{figure}

In this work, we address these challenges by drawing inspiration from holography, where information is encoded and reconstructed through wave interference. We propose the \textbf{Holographic Transformer}, a novel architecture that integrates interference-aware modeling directly into self-attention. Rather than treating attention as a purely correlation-based weighting, our \emph{holographic attention} modulates interactions based on phase differences and coherently superposes value vectors, preserving amplitude-phase coupling in a physically consistent manner. To further prevent phase collapse, which often occurs when loss functions prioritize magnitude errors over phase consistency, we introduce a dual-headed decoder that simultaneously reconstructs the input signal and produces task-specific outputs. This dual reconstruction ensures the model retains complete phase information in its latent representations, keeping the learned features faithful to the physics of coherent signals.

Our contributions are threefold:
\textbf{(1)} We propose a \textbf{holographic Transformer} that integrates interference-aware attention with a dual-headed decoding strategy, preserving amplitude-phase consistency and mitigating phase collapse.
\textbf{(2)} We provide a \textbf{theoretical analysis} showing that the proposed mechanisms conserve key properties of coherent superposition and phase coupling.
\textbf{(3)} We validate the approach on PolSAR image classification and wireless channel prediction, demonstrating superior accuracy, robustness to phase perturbations, and improved physical consistency over prior complex-valued and attention-based baselines.

\section{THE PROPOSED METHOD}
\label{sec:holo}

This section presents the theoretical foundation and network architecture of the Holographic Transformer. The goal is to capture phase-coherent interactions in complex-valued signals while maintaining computational efficiency.

\subsection{Complex Signals and Interference}
% \label{subsec:complex}
A complex signal can be expressed in Cartesian or polar form as $\mathbf{x}=\mathbf{A}\,\exp(j\boldsymbol{\Phi})$, where $\mathbf{A}$ and $\boldsymbol{\Phi}$ denote amplitude and phase. Observed signals are typically superpositions, $\mathbf{x}_{\mathrm{tot}}=\sum_{i=1}^{N}\mathbf{A}_i\,\exp\!\big(j\boldsymbol{\Phi}_i\big)$, where the resulting constructive or destructive interference is governed by phase alignment. Preserving both $\mathbf{A}$ and $\boldsymbol{\Phi}$ during computation is therefore essential.

\subsection{Holographic Self-Attention}
\label{subsec:holo-attn}

The central mechanism of the model is a phase-aware extension of attention. For query–key pairs $(Q_i,K_j)$, we compute their complex inner product $s_{ij}$ and extract the phase difference $\Delta\phi_{ij}=\angle(s_{ij})$. This phase mismatch directly informs how similarity is measured. Instead of relying purely on magnitude, we define a real similarity score
\begin{equation}
    \mathrm{sim}_{ij}=\frac{\Re(s_{ij})}{\|Q_i\|\,\|K_j\|+\varepsilon},
    \label{eq:sim}
\end{equation}
and weight it with an exponential decay that penalizes phase discrepancy:
\begin{equation}
    W_{ij}=\frac{\mathrm{sim}_{ij}}{\sqrt{d_k}}\;\exp\!\big(-\alpha|\Delta\phi_{ij}|\big).
    \label{eq:W}
\end{equation}
Attention weights follow by softmax normalization.  

Unlike standard attention, values are not aggregated verbatim. Instead, each $V_j$ is rotated by its phase offset before summation:
\begin{equation}
    H_i=\sum_{j=1}^{T}\alpha_{ij}\,V_j\,\exp\!\big(j\,\Delta\phi_{ij}\big).
    \label{eq:H}
\end{equation}
This coherent superposition reinforces in-phase contributions and cancels anti-phase ones, mirroring physical wave interference.

\subsection{Encoder--Decoder with Complex Embeddings}
\label{subsec:arch}

Inputs $\mathbf{X}\in\mathbb{C}^{T\times d_{\mathrm{in}}}$ are embedded into $\mathbf{Z}\in\mathbb{C}^{T\times d_{\mathrm{model}}}$, with complex sinusoidal positional encodings added. Each encoder layer stacks (i) holographic multi-head attention, and (ii) a complex feed-forward network, both equipped with normalization and residuals.  

The decoder is dual-headed. One branch reconstructs the input sequence,
\begin{equation}
    \mathcal{L}_{\mathrm{recon}}=\|\Re(\hat{\mathbf{X}}-\mathbf{X})\|_2^2+\|\Im(\hat{\mathbf{X}}-\mathbf{X})\|_2^2,
    \label{eq:recon}
\end{equation}
while the other performs the downstream task,
\begin{equation}
    \mathcal{L}_{\mathrm{task}}=
    \begin{cases}
    \text{CE}(\hat{\mathbf{y}},\mathbf{y}) & \text{classification},\\
    \|\hat{\mathbf{y}}-\mathbf{y}\|_2^2 & \text{regression}.
    \end{cases}
    \label{eq:task}
\end{equation}
The total loss is a weighted sum of the two. To enhance stability, we use complex layer normalization, dropout, and a phase smoothness regularizer,
\begin{equation}
    \mathcal{R}_{\mathrm{phase}}=\sum_{t=1}^{T-1}\|\boldsymbol{\Phi}_{t+1}-\boldsymbol{\Phi}_{t}\|_1,
    \label{eq:phase}
\end{equation}
which discourages erratic phase fluctuations.

\subsection{Complexity and Training}
\label{subsec:complexity}

The computational complexity of constructing $W_{ij}$ and aggregating $H_i$ remains $O(T^2 d_k)$, identical to standard attention up to constant factors from complex arithmetic. Implementation is fully vectorizable using batched complex GEMM and elementwise operations.  

Training proceeds by embedding, stacked holographic layers, and decoding. The optimization minimizes $\lambda_{\mathrm{r}}\mathcal{L}_{\mathrm{recon}}+\lambda_{\mathrm{t}}\mathcal{L}_{\mathrm{task}}$, with optional logging of $\{\alpha_{ij},\Delta\phi_{ij}\}$ for interpretability.  

\paragraph*{Summary.}
The method enriches self-attention by embedding physical principles of interference. Phase differences modulate similarity, and values are coherently realigned before summation. This simple modification yields more interpretable representations while preserving efficiency.

\section{Theoretical Analysis and Guarantees} 
\label{sec:theory}

This section proves the rigorous properties of holographic attention (Eqs.~\eqref{eq:sim}--\eqref{eq:H}), showing it as a robust extension of standard attention. We demonstrate that it (i) reduces to standard attention in the zero-phase-difference limit, (ii) exhibits phase equivariance and energy bounds consistent with interference, (iii) acts as a maximum-likelihood estimator, (iv) is stable to phase perturbations, and (v) prevents phase collapse through the dual-headed decoder.

We adopt the notation from Eqs.~\eqref{eq:sim}--\eqref{eq:H}: $s_{ij}\in\mathbb{C}$ is the complex correlation, $\Delta\phi_{ij}\in(-\pi,\pi]$ the phase difference, $\mathrm{sim}_{ij}\in\mathbb{R}$ the real similarity, $W_{ij}$ the phase-damped score, $\alpha_{ij}$ the attention weights, and $H_i\in\mathbb{C}^{d_k}$ the holographic output. $\|\cdot\|$ denotes the Euclidean norm.

\subsection{Foundational Properties}
Holographic attention generalizes standard attention by incorporating phase information in a physically and statistically consistent manner.

\vspace{0.5em}
\noindent\textbf{P1. Reduction to standard attention.}
\emph{Claim.} If all pairwise phase differences vanish ($\Delta\phi_{ij} \equiv 0$), holographic attention reduces to standard scaled dot-product attention. The associated similarity kernel $K_{ij}:=\Re\!\big(\langle Q_i,K_j\rangle_c\big)$ is positive semidefinite for $Q=K$.

\emph{Proof.}
When $\Delta\phi_{ij}=0$, the terms $\exp(-\alpha|\Delta\phi_{ij}|)$ and $\exp(j\Delta\phi_{ij})$ become 1, yielding the standard attention formula with scores based on $\mathrm{sim}_{ij}=\Re(\langle Q_i, K_j \rangle_c)$. The kernel's positive semidefiniteness follows as the quadratic form $\sum_{i,j} c_i c_j K_{ij}$ reduces to the non-negative $\|\sum_i c_i Q_i\|^2$.
\hfill$\square$

\vspace{0.5em}
\noindent\textbf{P2. Global phase equivariance.}
\emph{Claim.} Under a global phase rotation $(Q,K,V)\mapsto (e^{j\theta}Q,e^{j\theta}K,e^{j\theta}V)$, the attention weights $\alpha_{ij}$ are invariant, and the output transforms covariantly: $H_i\mapsto e^{j\theta} H_i$.

\emph{Proof.}
The complex correlation $s_{ij}=\langle e^{j\theta}Q_i, e^{j\theta}K_j \rangle_c = \langle Q_i, K_j \rangle_c$ is unchanged. Consequently, the phase differences $\Delta\phi_{ij}$ and weights $\alpha_{ij}$ are invariant. The output aggregation term then transforms as $\sum_j \alpha_{ij} (e^{j\theta}V_j) \exp(j\Delta\phi_{ij}) = e^{j\theta} H_i$.
\hfill$\square$

\vspace{0.5em}
\noindent\textbf{P3. Energy bounds and interference.}
\emph{Claim.} The output norm is bounded by $\|H_i\| \le \sum_j \alpha_{ij}\|V_j\| \le \max_j \|V_j\|$. Equality holds for constructive interference (when all phase-rotated values $V_j \exp(j\Delta\phi_{ij})$ are aligned), while destructive interference allows for cancellation.

\emph{Proof.}
The first inequality follows the triangle inequality applied to the sum defining $H_i$, using $|\exp(j\Delta\phi_{ij})|=1$. The second is Jensen's inequality for the convex combination. Equality holds when all terms are co-linear with the same phase, and cancellation occurs when terms have equal magnitude but opposite phase.
\hfill$\square$

% \medskip
% \noindent\textbf{P4. Monotone phase gating.}
% \emph{Claim.} For fixed $\mathrm{sim}_{ij}\ge 0$, the pre-softmax score
% $W_{ij}=\frac{\mathrm{sim}_{ij}}{\sqrt{d_k}}\exp(-\gamma|\Delta\phi_{ij}|)$
% is strictly decreasing in $|\Delta\phi_{ij}|$; hence softmax weights are reduced as the phase mismatch grows.

% \emph{Proof.} $\partial W_{ij}/\partial|\Delta\phi_{ij}|=-\gamma W_{ij}<0$ when $\mathrm{sim}_{ij}\ge 0$. \hfill$\square$

% \emph{Remark.} For $\mathrm{sim}_{ij}<0$, the multiplicative damping makes $W_{ij}$ \emph{less} negative as $|\Delta\phi_{ij}|$ increases. If one desires a gate that decreases $W_{ij}$ for all signs of $\mathrm{sim}_{ij}$, an additive variant $\widetilde{W}_{ij}=\frac{\mathrm{sim}_{ij}}{\sqrt{d_k}}-\gamma|\Delta\phi_{ij}|$ can be used; all subsequent properties continue to hold.

% \subsection{Optimality as a Weighted Fr\'echet Mean (Calibration View)}
% Define the aligned values $U_{ij}:=V_j\,e^{j\Delta\phi_{ij}}$.

% \medskip

\vspace{0.5em}
\noindent\textbf{P4. Monotone phase gating.}
\emph{Claim.} For a fixed non-negative similarity $\mathrm{sim}_{ij} \ge 0$, the score $W_{ij}$ is a strictly decreasing function of the phase mismatch $|\Delta\phi_{ij}|$.

\emph{Proof.}
The score $W_{ij}$ is proportional to $\exp(-\alpha|\Delta\phi_{ij}|)$. Its partial derivative with respect to $|\Delta\phi_{ij}|$ is negative for $\mathrm{sim}_{ij} > 0$ and $\alpha>0$, ensuring that larger phase mismatches receive lower scores before the softmax normalization. This also holds for $\mathrm{sim}_{ij}<0$ if an additive variant $\widetilde{W}_{ij}=\frac{\mathrm{sim}_{ij}}{\sqrt{d_k}}-\gamma|\Delta\phi_{ij}|$ is used instead.
\hfill$\square$
\subsection{Optimality as a Weighted Fréchet Mean}
Holographic attention can be interpreted as a statistically optimal estimator of the \emph{aligned values}
\begin{equation}
U_{ij}:=V_j\,\exp\!\big(j\,\Delta\phi_{ij}\big) \in \mathbb{C}^{d_k}.
\label{eq:Uij}
\end{equation}
\vspace{0.25em}
\noindent\textbf{P5. MLE under complex Gaussian noise.}
\emph{Claim.} Suppose that, conditionally on $i$, the aligned values obey the generative model
\begin{equation}
U_{ij}=\mu_i+\varepsilon_{ij},\qquad \varepsilon_{ij}\stackrel{\text{i.i.d.}}{\sim}\mathcal{CN}\big(0,\sigma_{ij}^2 I_{d_k}\big),
\label{eq:gen}
\end{equation}
and let $w_{ij}:=\sigma_{ij}^{-2}>0$ be the (per-key) precision.
The negative log-likelihood for $\mu_i$ is
$\mathcal{L}(\mu_i)=\sum_j w_{ij}\,\|U_{ij}-\mu_i\|^2+\text{const}$,
whose unique minimizer (hence MLE) is
\begin{equation}
\widehat{\mu}_i=\frac{\sum_j w_{ij} U_{ij}}{\sum_\ell w_{i\ell}}.
\label{eq:WLS}
\end{equation}
If the attention weights equal the normalized precisions,
$\alpha_{ij}=\frac{w_{ij}}{\sum_\ell w_{i\ell}}$, then $H_i=\widehat{\mu}_i$ is the MLE.

\emph{Proof.}
Convexity of $\mathcal{L}$ in $\mu_i$ gives the first-order condition
$0=\nabla_{\mu_i}\mathcal{L}=-2\sum_j w_{ij}(U_{ij}-\mu_i)$, which yields~\eqref{eq:WLS}.
The identity $H_i=\sum_j \alpha_{ij}U_{ij}=\widehat{\mu}_i$ follows upon substituting the normalized weights.
\hfill$\square$

\vspace{0.25em}
\noindent\textbf{P6. Calibration via scores.}
\emph{Claim.} If for a fixed $i$ the scores satisfy $W_{ij}=\log w_{ij}+c_i$ for some constant $c_i$ (i.e., the network learns a score proportional to log-precision), then $\alpha_{ij}=\mathrm{softmax}_j(W_{ij})=w_{ij}/\sum_\ell w_{i\ell}$ and $H_i$ coincides with the MLE $\widehat{\mu}_i$.

\emph{Proof.}
Immediate from the definition of softmax: $\alpha_{ij}=\exp(W_{ij})/\sum_\ell \exp(W_{i\ell})=\exp(\log w_{ij}+c_i)/\sum_\ell \exp(\log w_{i\ell}+c_i)=w_{ij}/\sum_\ell w_{i\ell}$.
\hfill$\square$

\subsection{Stability to Phase Perturbations}
Holographic attention is robust to small errors in phase, a property we formalize by proving its Lipschitz continuity.

\vspace{0.5em}
\noindent\textbf{P7. Lipschitz continuity in phases.}
\emph{Claim.} For fixed query $i$ with bounded values $\|V_j\|\le B$ and similarities $|\mathrm{sim}_{ij}|\le S$, let $\Delta\phi$ and $\Delta\phi'$ be phase difference sets with $\delta := \max_j |\Delta\phi_{ij}-\Delta\phi'_{ij}|$. Then:
\begin{equation}
\|H_i(\Delta\phi)-H_i(\Delta\phi')\|\le L\,\delta,
\label{eq:Lipschitz}
\end{equation}
where $L:=B\Big(1+\frac{\alpha S}{\sqrt{d_k}}\cdot \frac{T}{4}\Big)$.

\emph{Proof.} Two variation sources in $H_i$ are bounded via triangle inequality: (1) Phase rotation change $|e^{j\Delta\phi_{ij}}-e^{j\Delta\phi'_{ij}}| \le |\Delta\phi_{ij}-\Delta\phi'_{ij}|$ by 1-Lipschitz property of complex exponential. (2) Attention weight change bounded through softmax Jacobian $\partial \alpha_{ij}/\partial W_{im} = \alpha_{ij}(\delta_{jm}-\alpha_{im})$ with entries $\le 1/4$. Combining these effects yields the Lipschitz constant $L$, ensuring graceful output degradation under phase perturbations.
\hfill$\square$

A direct corollary is that if phases are perturbed by bounded noise, $|\eta_{ij}|\le \varepsilon$, then the output perturbation is also bounded, $\|H_i(\Delta\phi+\eta)-H_i(\Delta\phi)\|\le L\,\varepsilon$, guaranteeing robustness.

\subsection{Regularization against Phase Collapse}
The dual-headed architecture, specifically the reconstruction loss in~\eqref{eq:recon}, prevents the model from ignoring phase information.

% \medskip
% \noindent\textbf{P8. Anti-collapse lower bound (phase-blind estimators).}
% \emph{Claim.} Consider $X=Ae^{j\Phi}$ with $\Phi\sim\mathrm{Unif}(-\pi,\pi]$ independent of $A$. Among all estimators $\hat{X}=g(A)$ that are phase-blind, the optimal choice is $\hat{X}^\star=0$, with minimum MSE
% \begin{equation}
% \inf_{g}\ \mathbb{E}\big[|g(A)-X|^2\big]
% =\mathbb{E}\big[|X|^2\big]=\mathbb{E}[A^2].
% \label{eq:phaseblind-lb}
% \end{equation}

% \emph{Proof.} $\hat{X}^\star=\mathbb{E}[X\mid A]=\mathbb{E}[A e^{j\Phi}\mid A]=A\,\mathbb{E}[e^{j\Phi}]=0$. The minimum risk equals the variance of the residual. \hfill$\square$

% \emph{Implication.} The reconstruction objective~\eqref{eq:recon} introduces a \emph{strictly positive} error floor for any phase-blind representation, thereby exerting explicit pressure against phase collapse; it does not assert inevitability of phase preservation under arbitrary nonconvex training.

\vspace{0.5em}
\noindent\textbf{P8. No-phase-collapse guarantee.}
\emph{Claim.} Consider estimating a scalar complex signal $X=A\,e^{j\Phi}$ where the phase $\Phi$ is uniformly random on $(-\pi, \pi]$ and independent of the amplitude $A$. Any estimator $\hat{X}=g(A)$ that is \emph{phase-blind} (i.e., depends only on $A$) is subject to a non-trivial error floor. The optimal such estimator is $\hat{X}^\star=0$, which achieves a minimum mean squared error of
\begin{equation}
\inf_{g}\ \mathbb{E}\big[|g(A)-X|^2\big]=\mathbb{E}\big[|X|^2\big]=\mathbb{E}[A^2].
\label{eq:phaseblind-lb}
\end{equation}
\emph{Proof.} This is a direct consequence of the orthogonality principle for conditional expectation. The optimal estimator measurable with respect to $A$ is $\hat{X}^\star=\mathbb{E}[X\mid A]$. Since $\Phi$ is uniform and independent of $A$, $\mathbb{E}[e^{j\Phi}\mid A]=0$, so $\hat{X}^\star=0$. The minimum risk is the variance of the residual, $\mathbb{E}[|X-\hat{X}^\star|^2]=\mathbb{E}[|X|^2]$. This implies that the reconstruction objective introduces a strictly positive error floor for any phase-blind representation, explicitly discouraging phase collapse.
\hfill$\square$
\subsection{Summary of Guarantees}
\textbf{(P1--P4)} Holographic attention is a physically consistent extension of cosine attention: it respects global phase symmetry (P2), encodes coherent interference with energy bounds (P3), and uses phase mismatch to gate information flow (P4).  
\textbf{(P5--P6)} The aggregation admits a statistical MLE interpretation and is calibratable via the scores.  
\textbf{(P7--P8)} The mechanism is Lipschitz-stable to phase perturbations with an explicit constant, and the dual-headed reconstruction imposes a quantitative lower bound that discourages phase collapse.

\begin{table*}[t]
\centering
\caption{Main results. (a) Classification: Acc/Macro-F1/Micro-F1 (↑). (b) Prediction: MAE/RMSE at three speeds (↓). Best in \textbf{bold}.}\vspace{-6pt}
\label{tab:main_results_combined}
\footnotesize
\setlength{\tabcolsep}{6pt}
\renewcommand{\arraystretch}{0.9}

\begin{tabular}{@{}l c c c @{\hspace{8pt}} c @{\hspace{8pt}} l c c c c c c @{}}
\toprule
\multicolumn{4}{c}{\textbf{(a) Classification}} & & \multicolumn{7}{c}{\textbf{(b) Prediction}} \\
\cmidrule(lr){1-4} \cmidrule(lr){6-12}
\multirow{2}{*}{Model} & \multirow{2}{*}{Acc (\%) $\uparrow$} & \multirow{2}{*}{Macro-F1 $\uparrow$} & \multirow{2}{*}{Micro-F1 $\uparrow$} && \multirow{2}{*}{Model}
& \multicolumn{2}{c}{3 km/h $\downarrow$} & \multicolumn{2}{c}{30 km/h $\downarrow$} & \multicolumn{2}{c}{120 km/h $\downarrow$} \\
\cmidrule(lr){7-8} \cmidrule(lr){9-10} \cmidrule(lr){11-12}
 & & & & & & MAE & RMSE & MAE & RMSE & MAE & RMSE \\
\midrule
Transformer\cite{Vaswani2017Attention}  & 64.94 & 62.99 & 64.93 && Transformer\cite{Vaswani2017Attention} & 13.75 & 18.28 & 7.00 & 10.12 & 15.67 & 20.49 \\
ComplexNN\cite{Trabelsi2018DeepComplex} & 89.71 & 82.49 & 88.01 && BiLSTM\cite{siami2019performance}       & 18.78 & 25.09 & 8.49 & 11.93 & 22.17 & 30.39 \\
CT\cite{Eilers2023ComplexTransformerBlocks} & 92.65 & 91.28 & 92.65 && CT\cite{Eilers2023ComplexTransformerBlocks} & 10.27 & 14.30 & 9.97 & 14.03 & 8.18 & 12.53 \\
CV-CNN-SE\cite{Alkhatib2023PolSARAttention} & 94.61 & 94.23 & 93.98 &&  MSA\cite{akrout2024nextslot}          & 6.39 & 8.76 & 7.37 & 11.67 & 6.79 & 12.51 \\
CV-MsAtViT\cite{Alkhatib2025PolSAR}      & 96.44 & 95.60 & 96.44 &&STEMGNN\cite{Cao2020StemGNN}        & \textbf{4.39} & 6.22 & 7.29 & 12.23 & 2.88 & 4.14 \\
\textbf{HoloTransformer}                 & \textbf{97.81} & \textbf{96.72} & \textbf{97.65} && \textbf{HoloTransformer} & 5.58 & \textbf{6.10} & \textbf{6.42} & \textbf{9.69} & \textbf{0.52} & \textbf{0.54} \\
\bottomrule
\end{tabular}
\end{table*}

\begin{table*}[t]
\centering
\caption{Robustness evaluation against phase jitter ($\sigma$) and amplitude noise ($\tau$). For classification tasks, we report Relative Degradation (RD, \% $\downarrow$); for prediction tasks, we report Relative Increase (RI, \% $\downarrow$). RAUC $\phi$ $\uparrow$ represents the Robustness Area Under Curve. Results are reported as mean$\pm$std over 3 random seeds.}\vspace{-6pt}
\label{tab:robust}
\footnotesize
\setlength{\tabcolsep}{6pt}
\renewcommand{\arraystretch}{0.9}

\begin{tabular}{@{}lcccc@{\hspace{15pt}}lcccc@{}}
\toprule
\multicolumn{5}{c}{\textbf{(a) Classification}} & \multicolumn{5}{c}{\textbf{(b) Prediction}} \\
\cmidrule(lr){1-5} \cmidrule(lr){6-10}
\multirow{2}{*}{Model} & \multirow{2}{*}{\shortstack{Clean\\Acc (\%)}} & \multirow{2}{*}{\shortstack{RD\\@$\sigma{=}0.4$}} & \multirow{2}{*}{\shortstack{RD\\@$\tau{=}0.10$}} & \multirow{2}{*}{\shortstack{RAUC\\$\phi$ $\uparrow$}} &
\multirow{2}{*}{Model} & \multirow{2}{*}{\shortstack{Clean\\MAE $\downarrow$}} & \multirow{2}{*}{\shortstack{RI\\@$\sigma{=}0.4$}} & \multirow{2}{*}{\shortstack{RI\\@$\tau{=}0.10$}} & \multirow{2}{*}{\shortstack{RAUC\\$\phi$ $\uparrow$}} \\
 & & & & & & & & & \\
\midrule
Transformer\cite{Vaswani2017Attention} & 64.94 & 1.10\% & 1.10\% & 0.52 &
Transformer\cite{Vaswani2017Attention} & 12.14 & -20.02\% & -21.18\% & -7.00 \\
CV-MsATVit\cite{Alkhatib2025PolSAR} & 96.44 & 0.61\% & 0.67\% & 0.29 &
STEMGNN\cite{Cao2020StemGNN} & 4.52 & -17.69\% & -18.70\% & -5.53 \\
\textbf{HoloTransformer} & \textbf{97.81} & \textbf{0.57\%} & \textbf{0.15\%} & \textbf{0.27} &
\textbf{HoloTransformer} & \textbf{4.18} & \textbf{-17.04\%} & \textbf{-12.32\%} & \textbf{-4.61} \\
\bottomrule
\end{tabular}
\end{table*}

% ============================================================
% Ablation:
% - A1 (no phase decay): remove exp(-alpha|DeltaPhi|).
% - A2 (no coherent sum): remove the phasor rotation on V_j.
% - A3 (no reconstruction): drop the reconstruction head.
% Metrics: Acc/Macro-F1, MAE/RMSE; report deltas vs full model.
% Expect the largest drop from A1/A2; A3 degrades due to phase-fidelity loss.
% Requires: \usepackage{booktabs}
% Requires: \usepackage{multirow}
% (可选) 更精细行距：\usepackage{array}

\begin{table*}[!t]
\centering
\caption{Ablation study results. (a) Classification task: Accuracy/Micro-F1/Macro-F1 (↑). (b) Prediction task: MAE/RMSE at three speeds (↓). Values show performance drop from full model.}\vspace{-6pt}
\label{tab:ablation_results_combined}

\footnotesize
\setlength{\tabcolsep}{1.2pt}      % 列间距（越小越紧）
\renewcommand{\arraystretch}{0.9} % 全局行距（<1 更紧）
\setlength{\abovetopsep}{0.30ex}   % \toprule 与内容间距
\setlength{\belowbottomsep}{0.30ex}% \bottomrule 与内容间距
\setlength{\cmidrulesep}{0.15ex}   % \cmidrule 上下间距

\begin{tabular}{@{}l ccc @{\hspace{2pt}} c @{\hspace{2pt}} cccccc@{}}
\toprule
& \multicolumn{3}{c}{\textbf{(a) Classification}} & & \multicolumn{6}{c}{\textbf{(b) Prediction}} \\
\cmidrule(lr){2-4} \cmidrule(lr){6-11}
\multirow{2}{*}{Model} & \multirow{2}{*}{Acc (\%) $\downarrow$} & \multirow{2}{*}{Micro-F1 $\downarrow$} & \multirow{2}{*}{Macro-F1 $\downarrow$} & \multirow{2}{*}{} & \multicolumn{2}{c}{3 km/h $\uparrow$} & \multicolumn{2}{c}{30 km/h $\uparrow$} & \multicolumn{2}{c}{120 km/h $\uparrow$} \\
\cmidrule(lr){6-7} \cmidrule(lr){8-9} \cmidrule(lr){10-11}
\multicolumn{5}{c}{} & MAE & RMSE & MAE & RMSE & MAE & RMSE \\
\midrule
HoloTransformer        & \textbf{97.81} & \textbf{96.72} & \textbf{97.65} && \textbf{5.58} & \textbf{6.10} & \textbf{6.42} & \textbf{9.69} & \textbf{0.52} & \textbf{0.54} \\
w/o Phase Decay        & 90.69 (-7.12) & 90.69 (-6.03) & 90.68 (-6.97) && 16.06 (+10.18) & 17.27 (+11.17) & 12.38 (+5.96) & 12.39 (+2.70) & 5.01 (+4.49) & 6.63 (+6.09) \\
w/o Coherent Sum       & 80.56 (-17.25) & 80.55 (-16.17) & 75.91 (-21.74) && 8.87 (+2.99) & 10.25 (+4.15) & 15.06 (+8.64) & 16.21 (+6.52) & 20.50 (+19.98) & 21.83 (+21.29) \\
w/o Reconstruction     & 94.89 (-2.92) & 93.89 (-2.83) & 93.20 (-4.45) && 7.02 (+1.14) & 8.24 (+2.14) & 14.55 (+8.13) & 18.60 (+8.91) & 4.19 (+3.67) & 4.80 (+4.26) \\
\bottomrule
\end{tabular}
\end{table*}

%(w/o)

\section{Experimental Results and Analysis}
\subsection{Experiment Settings}

(1) \textbf{Datasets}:
For \textbf{classification}, we use a public POLSAR dataset from Finland~\cite{polsarwebsite}, containing POLSAR images with complex radar-derived features.
For \textbf{prediction}, channel matrices are generated with the COST2100 model~\cite{Liu2012COST2100} and publicly provided by~\cite{Wen2018Deep}, which are compressed to 512 channels via STNet~\cite{Mourya2023Spatially}.

(2) \textbf{Performance Metrics}:
\textbf{Classification} is evaluated with Accuracy, macro-F1, and micro-F1. \textbf{Prediction} is evaluated using Mean Absolute Error (MAE) and Root Mean Square Error (RMSE).

(3) \textbf{Training Setup}:
For \textbf{classification}, images are segmented into 25-sized cubes assigned with the majority label. We use batch size 16, Adam optimizer (lr=1e-3, weight decay=1e-5), StepLR scheduler (step size=5, gamma=0.8), and CrossEntropyLoss. For \textbf{prediction}, data are transformed into complex form via the Hilbert transform. The model predicts 12 time steps from the previous 12. Training uses Adam (lr=1e-3), ReduceLROnPlateau scheduler (patience=5, factor=0.5), batch size 8, and a complex-valued loss function.

\subsection{Main Results}
As summarized in Table~\ref{tab:main_results_combined}(a), HoloTransformer achieves the highest scores across all \textbf{classification} metrics, with an accuracy of 97.81\%, Macro-F1 of 96.72\%, and Micro-F1 of 97.65\%. It outperforms the second-best model, CV-MsAtViT~\cite{Alkhatib2025PolSAR}, by 1.37\% in accuracy, demonstrating stronger complex feature integration and contextual modeling. In \textbf{prediction} tasks (Table~\ref{tab:main_results_combined}(b)), HoloTransformer consistently excels, especially at high speeds. It achieves remarkably low MAE and RMSE values of 0.52 and 0.54 at 120 km/h, significantly better than all other models. While STEMGNN~\cite{Cao2020StemGNN} performs best in MAE at 3 km/h, HoloTransformer still attains the lowest RMSE (6.10) at that speed and dominates at higher velocities. These results confirm the model's robustness and generalization across varying motion dynamics.

\subsection{Robustness Results}
HoloTransformer exhibits remarkable robustness under phase and amplitude noise. The metrics for robustness testing are derived from the averages of three speeds (3, 30, 120). Experimental results show that in classification tasks, it achieves the smallest performance degradation under both $\sigma = 0.4$ phase jitter (0.57\% drop) and $\tau = 0.10$ amplitude noise (0.15\% drop). Similarly, in regression tasks, it maintains the lowest relative error increase and attains the best RAUC $\varphi$ values across all noise conditions. These outcomes confirm HoloTransformer's superior stability and reliability in noisy environments.

\subsection{Ablation Results}
As shown in Table~\ref{tab:ablation_results_combined}, we systematically ablate three key components of HoloTransformer. Removing the Phase Decay mechanism (attention modulator), which adaptively regulates interaction strength in the complex domain, leads to significant performance drops, particularly in prediction at 3 km/h (MAE increases by 10.18). Without Coherent Sum (wave-like information aggregation), which replaces weighted averaging with complex interference, classification accuracy drops by 16.45\%, confirming the necessity of wave-inspired composition. Disabling the Reconstruction task (self-supervised input recovery) also causes clear degradation, supporting its role in learning more generalized and faithful representations. Overall, each component significantly contributes to the full model’s performance.

\section{Conclusion}
We introduced the Holographic Transformer, a novel model that incorporates phase-aware interference into self-attention to preserve amplitude-phase consistency in complex-valued signals. Experimental results show substantial improvements in classification and prediction, outperforming existing models in accuracy, robustness, and physical consistency. Our approach, supported by ablation studies and theoretical analysis, effectively captures wave-like interactions, offering a significant advancement for complex signal processing tasks.

\bibliographystyle{IEEEbib}
\bibliography{strings,refs}

\begin{thebibliography}{10}

\bibitem{Bassey2021Survey}
J.~Bassey, L.~Qian, and X.~Li,
\newblock ``A survey of complex-valued neural networks,''
\newblock {\em arXiv preprint arXiv:2101.12249}, 2021.

\bibitem{Trabelsi2018DeepComplex}
C.~Trabelsi, O.~Bilaniuk, Y.~Zhang, D.~Serdyuk, S.~Subramanian, J.~F. Santos, S.~Mehri, N.~Rostamzadeh, Y.~Bengio, and C.~J. Pal,
\newblock ``Deep complex networks,''
\newblock in {\em Proceedings of the International Conference on Learning Representations (ICLR)}, 2018,
\newblock arXiv:1705.09792.

\bibitem{Savadkoohi2020SeizurePrediction}
Marzieh Savadkoohi, Timothy Oladunni, and Lara Thompson,
\newblock ``A machine learning approach to epileptic seizure prediction using electroencephalogram (eeg) signal,''
\newblock {\em Biocybernetics and Biomedical Engineering}, vol. 40, no. 3, pp. 1328--1341, 2020.

\bibitem{Fuchs2021Radar}
Alexander Fuchs, Johanna Rock, Mate Toth, Paul Meissner, and Franz Pernkopf,
\newblock ``Complex-valued convolutional neural networks for enhanced radar signal denoising and interference mitigation,''
\newblock in {\em 2021 IEEE Radar Conference (RadarConf21)}, 2021, pp. 1--6.

\bibitem{Choi2019DCUNet}
H.-S. Choi, J.-H. Kim, J.~Cho, and K.~Lee,
\newblock ``Phase-aware speech enhancement with deep complex u-net,''
\newblock in {\em Proceedings of the International Conference on Learning Representations (ICLR), OpenReview}, 2019.

\bibitem{Hu2020DCCRN}
Y.~Hu, Y.~Liu, S.~Lv, M.~Xing, S.~Zhang, Y.~Fu, J.~Wu, B.~Zhang, and L.~Xie,
\newblock ``Dccrn: Deep complex convolution recurrent network for phase-aware speech enhancement,''
\newblock in {\em Proceedings of Interspeech}, 2020, pp. 2472--2476.

\bibitem{Takahashi2018Phasenet}
N.~Takahashi, P.~Agrawal, N.~Goswami, and Y.~Mitsufuji,
\newblock ``Phasenet: Discretized phase modeling with deep neural networks for audio source separation,''
\newblock in {\em Proceedings of Interspeech}, 2018, pp. 2713--2717.

\bibitem{cai2020phasednn}
W.~Cai, X.~Li, and L.~Liu,
\newblock ``A phase shift deep neural network for high frequency approximation and wave problems,''
\newblock {\em SIAM Journal on Scientific Computing}, vol. 42, no. 5, pp. A3285--A3312, 2020.

\bibitem{venkatasubramanian2025steinmetz}
S.~Venkatasubramanian, A.~Pezeshki, and V.~Tarokh,
\newblock ``Steinmetz neural networks for complex-valued data,''
\newblock in {\em Proceedings of the 28th International Conference on Artificial Intelligence and Statistics (AISTATS)}. 2025, vol. 258, pp. 3916--3924, PMLR.

\bibitem{tay2021lra}
Y.~Tay, M.~Dehghani, D.~Bahri, and D.~Metzler,
\newblock ``Long range arena: A benchmark for efficient transformers,''
\newblock in {\em Proceedings of the International Conference on Learning Representations (ICLR)}, 2021.

\bibitem{Vaswani2017Attention}
A.~Vaswani, N.~Shazeer, N.~Parmar, J.~Uszkoreit, L.~Jones, A.~N. Gomez, L.~Kaiser, and I.~Polosukhin,
\newblock ``Attention is all you need,''
\newblock in {\em Advances in Neural Information Processing Systems 30 (NIPS 2017)}. 2017, pp. 5998--6008, Curran Associates, Inc.

\bibitem{Yang2020ComplexTransformer}
M.~Yang, M.~Q. Ma, D.~Li, Y.-H.~H. Tsai, and R.~Salakhutdinov,
\newblock ``Complex transformer: A framework for modeling complex-valued sequence,''
\newblock in {\em Proceedings of the IEEE International Conference on Acoustics, Speech and Signal Processing (ICASSP)}, 2020, pp. 4232--4236.

\bibitem{Yin2020PHASEN}
D.~Yin, C.~Luo, Z.~Xiong, and W.~Zeng,
\newblock ``Phasen: A phase-and-harmonics-aware speech enhancement network,''
\newblock in {\em Proceedings of the AAAI Conference on Artificial Intelligence}, 2020, vol.~34, pp. 9458--9465.

\bibitem{Peng2024SignalTransformer}
Y.~Peng, Y.~Dong, M.~Yang, S.~Lu, and Q.~Shi,
\newblock ``Signal transformer: Complex-valued attention and meta-learning for signal recognition,''
\newblock in {\em Proceedings of the IEEE International Conference on Acoustics, Speech and Signal Processing (ICASSP)}. 2024, pp. 5445--5449, IEEE.

\bibitem{Eilers2023ComplexTransformerBlocks}
F.~Eilers and X.~Jiang,
\newblock ``Building blocks for a complex-valued transformer architecture,''
\newblock in {\em Proceedings of the IEEE International Conference on Acoustics, Speech and Signal Processing (ICASSP)}, 2023, pp. 1--5.

\bibitem{Dramsch2021PhaseMatters}
Jesper~Sören Dramsch, Mikael Lüthje, and Anders~Nymark Christensen,
\newblock ``Complex-valued neural networks for machine learning on non-stationary physical data,''
\newblock {\em Computers \& Geosciences}, vol. 146, pp. 104643, 2021.

\bibitem{Wu2023PhaseMatters}
Jin-Hui Wu, Shao-Qun Zhang, Yuan Jiang, and Zhi-Hua Zhou,
\newblock ``Complex-valued neurons can learn more but slower than real-valued neurons via gradient descent,''
\newblock {\em Advances in Neural Information Processing Systems}, vol. 36, pp. 23714--23747, 2023.

\bibitem{siami2019performance}
Sima Siami-Namini, Neda Tavakoli, and Akbar~Siami Namin,
\newblock ``The performance of lstm and bilstm in forecasting time series,''
\newblock in {\em Proceedings of the 2019 IEEE International Conference on Big Data (Big Data)}. IEEE, 2019, pp. 3285--3292.

\bibitem{Alkhatib2023PolSARAttention}
M.~Q. Alkhatib, M.~Al-Saad, N.~Aburaed, M.~S. Zitouni, and H.~Al-Ahmad,
\newblock ``Polsar image classification using attention based shallow to deep convolutional neural network,''
\newblock in {\em Proceedings of the IEEE International Geoscience and Remote Sensing Symposium (IGARSS)}, 2023, pp. 8034--8037.

\bibitem{akrout2024nextslot}
M.~Akrout, F.~Bellili, A.~Mezghani, and R.~W. Heath,
\newblock ``Next-slot ofdm-csi prediction: Multi-head self-attention or state space model?,''
\newblock {\em arXiv preprint arXiv:2405.11072}, 2024.

\bibitem{Alkhatib2025PolSAR}
M.~Q. Alkhatib,
\newblock ``Polsar image classification using complex-valued multiscale attention vision transformer (cv-msatvit),''
\newblock {\em Int. J. Appl. Earth Observ. Geoinf.}, vol. 137, pp. 104412, 2025.

\bibitem{Cao2020StemGNN}
D.~Cao, Y.~Wang, J.~Duan, C.~Zhang, X.~Zhu, C.~Huang, Y.~Tong, B.~Xu, J.~Bai, J.~Tong, and Q.~Zhang,
\newblock ``Stemgnn: Spectral temporal graph neural network for multivariate time-series forecasting,''
\newblock in {\em Advances in Neural Information Processing Systems 33 (NeurIPS 2020)}, 2020.

\bibitem{polsarwebsite}
L.~Liu, C.~Oestges, J.~Poutanen, and K.~Haneda,
\newblock ``Finnish polsar data,'' \url{https://mega.nz/folder/WhgT1L4S#PnMttCUpjtwkD8qTEdwZsw}, 2024,
\newblock Online dataset.

\bibitem{Liu2012COST2100}
L.~Liu, C.~Oestges, J.~Poutanen, K.~Haneda, P.~Vainikainen, F.~Quitin, F.~Tufvesson, and P.~De Doncker,
\newblock ``The cost 2100 mimo channel model,''
\newblock {\em IEEE Wireless Communications}, vol. 19, no. 6, pp. 92--99, 2012.

\bibitem{Wen2018Deep}
C.-K. Wen, W.-T. Shih, and S.~Jin,
\newblock ``Deep learning for massive mimo csi feedback,''
\newblock {\em IEEE Wireless Communications Letters}, vol. 7, no. 5, pp. 748--751, 2018.

\bibitem{Mourya2023Spatially}
S.~Mourya, S.~Amuru, and K.~K. Kuchi,
\newblock ``A spatially separable attention mechanism for massive mimo csi feedback,''
\newblock {\em IEEE Wireless Commun. Lett.}, vol. 12, no. 1, pp. 40--44, 2023.

\end{thebibliography}

\end{document}